\documentstyle[aps,prb,eqsecnum]{revtex}
\begin{document}
\title{Gaussian, exponential, and power-law decay of time-dependent
correlation functions in quantum spin chains}

\author{Joachim Stolze }

\address{Physikalisches Institut, Universit\"at Bayreuth, 95440
Bayreuth, Germany\\ and Institut f\"ur Physik, Universit\"at Dortmund, 44221
Dortmund, Germany \cite{byline} }

\author{Angela N\"oppert}

\address{Institut f\"ur Physik, Universit\"at Dortmund, 44221
Dortmund, Germany}

\author{Gerhard M\"uller}

\address{Department of Physics, The University of Rhode Island, Kingston, R.I.
02881-0817}

\date{\today}

\maketitle

\begin{abstract}
Dynamic spin correlation functions
$\langle S _i^x (t)S_j^x \rangle$ for the 1D
$S=1/2$ $XX$ model
$H = -J\Sigma_i \{S_i^x S_{i+1}^x + S_i^y S_{i+1}^y \}$
are calculated exactly for finite open chains with up to $N=10000$ spins.
Over a certain time range the results are free of finite-size effects
and thus represent correlation functions of an infinite chain (bulk regime)
or a semi-infinite chain (boundary regime).
In the bulk regime, the long-time asymptotic decay as inferred by
extrapolation is Gaussian at $T=\infty$, exponential at
$0 < T < \infty$, and power-law $(\sim t^{-1/2} )$ at $T=0$, in
agreement with exact results.
In the boundary regime, a power-law decay obtains at all temperatures;
the characteristic exponent is universal at $T=0$ $(\sim t^{-1} )$ and
at $0 < T < \infty$ $(\sim t^{-3/2} )$, but is site-dependent at
$T=\infty$.
In the high-temperature regime $(T/J \gg 1)$ and in the low-temperature
regime $(T/J \ll 1)$, crossovers between different decay laws can be
observed in $\langle S_i^x (t)S_j^x \rangle$.
Additional crossovers are found between bulk-type and boundary-type decay
for $i=j$ near the boundary, and between
space-like and time-like behavior for $i \neq j$.
\end{abstract}
\pacs{75.10.Jm,  75.40.Gb}
%
\section{Introduction}
\label{I}
The long-time behavior of correlations for quantum many-body systems in
general and for quantum spin systems in particular has been a notoriously
difficult subject of theoretical research.
Approximation schemes tend to have little reliability in this field.
There exist very few exact results for nontrivial cases, and many of them
exhibit non-generic features for one reason or another.
In classical many-body systems, the long-time correlations can be investigated
by means of computer simulations, but no practical quantum counterpart
of that approach exists.
\par
In some cases, useful conclusions on the long-time behavior can be drawn
from a moment expansion via rigorous bounds, but the time intervals over
which stringent bounds can be established are often too short for that
purpose.\cite{RMP86,BL92,BL93}
The continued-fraction analysis based on the same number of moments
can be used to predict the exponent of a power-law infrared singularity
in the frequency domain, but this approach tends
to be insensitive to subtle changes in the long-time decay if it does not
involve power laws.\cite{SVM92,BVSM94}
\par
The one-dimensional (1D) $S=1/2$ $XY$ model,
\begin{equation}
H_{XY} = - \sum_{i = 1}^{N-1}  \left\{ J_x S_{i}^{x}
S_{i + 1}^{x} + J_y S_{i}^{y} S_{i + 1}^{y} + hS_i^z \right\} \, ,
\label{I.1}
\end{equation}
is one of the very few many-body systems with nontrivial dynamics for which
time-dependent correlation functions have been calculated exactly
at zero \cite{N67,KHS70,MBA71,VT78,MPS83,MS84}
and nonzero \cite{SJL75,BJ76,CP77,IIKS93} temperatures.
This model is equivalent to a system of noninteracting lattice fermions.
\cite{LSM61,K62} The spin correlation function
$\langle S_i^z (t)S_j^z \rangle$ is a simple fermion
density correlation function, and the function
$\langle S_i^x (t)S_j^x \rangle$
can be reduced to a determinant whose size increases linearly with the
number of sites between $i$ or $j$ and the nearest boundary of the chain.
\par
The focus of this study is on the dynamics of the spin $x$-components
for the special case $J_x = J_y$, $h=0$ of (\ref{I.1}) -- the $XX$ model.
For quite some time it has been known that the function
$\langle S_i^x (t)S_i^x \rangle$ of the infinite system exhibits a Gaussian
decay at
$T=\infty$ \cite{SJL75,BJ76,CP77} and a power-law decay at $T=0$.
\cite{MBA71,VT78}
A more recent study states that the long-time asymptotic decay of the
same correlation function is exponential at finite nonzero temperatures.
\cite{IIKS93} Numerical evidence for exponential decay was
also found for $\langle S_i^x (t)S_i^x \rangle$ in the $XXZ$
model at $T = \infty$.\cite{BVSM94}
\par
For a semi-infinite $XX$ chain the available evidence indicates that the
function $\langle S_i^x (t)S_i^x \rangle$ exhibits a power-law decay
at all temperatures.
Rigorous results exist for $T=\infty$,\cite{SVM92,SMC91} and the result of
a finite-chain study for $T=0$,\cite{PM78} but no results for
$0<T<\infty$ appear to have existed prior to this study.
\par
The purpose of this paper is (i) to fill in the missing links on the question
of long-time asymptotic behavior and (ii) to elucidate various kinds of
crossovers between the different decay laws that can be found in the
autocorrelation and paircorrelation functions $\langle S_i^x (t)S_j^x \rangle$.
The determinantal expressions for these functions have been known for a
long time.\cite{MBA71} However, only with today's advanced computer
technology can they be evaluated for systems large enough to yield data
from which conclusions can be drawn with some confidence about the
long-time asymptotics for infinite and semi-infinite chains at arbitrary
temperatures.\cite{note0}
In Sec. II we describe the method used for our analysis, in Sec. III
we present our results for the infinite system, and in Sec. IV we discuss
boundary effects.
%
\section{Fermion representation}
\label{II}
The Jordan-Wigner transformation \cite{LSM61,K62}
\begin{equation}
S_i^z = a_i^{\dag} a_i - \frac{1}{2} \, ,
\label{II.1}
\end{equation}
\begin{equation}
S_i^+ = (-1)^{ \sum_{k=1}^{i-1} a_k^{\dag} a_k } a_i^{\dag} \, , \;\;\;
S_i^- = a_i (-1)^{ \sum_{k=1}^{i-1} a_k^{\dag} a_k } \, ,
\label{II.2}
\end{equation}
between the component and ladder operators $S_i^z$,
$S_i^{\pm} = S_i^x \pm iS_i^y$ for an array of localized spins with $S=1/2$
and the creation and annihilation operators $a_i^{\dag}$, $a_i$
of an array of fermions converts
the Hamiltonian of an open-ended $XX$ chain,
\begin{equation}
H_{XX} = -J \sum_{i = 1}^{N-1}  \left\{S_{i}^{x} S_{i + 1}^{x}
+ S_{i}^{y} S_{i + 1}^{y} \right\} \, ,
\label{II.3}
\end{equation}
into a Hamiltonian of noninteracting fermions
\begin{equation}
H_{XX} = - \frac{J}{2} \sum_{i=1}^{N-1} (a_i^{\dag} a_{i+1} +
a_{i+1}^{\dag} a_i) \, .
\label{II.4}
\end{equation}
The energies of the one-particle eigenstates are
\begin{equation}
\varepsilon_k = - J \cos k \; \; \;  ; \; \; \;  k = \frac {\nu \pi}{N+1} \; \;
 \;  ;  \; \; \; \nu = 1, \ldots, N \, .
\label{II.5}
\end{equation}
\par
The spin correlation functions $\langle S_i^z (t)S_j^z \rangle$ are
then in essence density
correlation functions for the band (\ref{II.5}) of free fermions.
Their characteristic $t^{-1}$ long-time asymptotic behavior at zero and
nonzero temperatures \cite{N67,KHS70,MS84} is a consequence
of the band-edge singularities in the one-particle
density of state and (for $T=0$) the singularity at $\omega=0$ generated
by the Fermi function.
For the spins at the boundary of a semi-infinite chain, different power-law
decays of $\langle S_i^z (t)S_j^z \rangle$ pertain to $T=0$
($\sim t^{-2}$ if both $i$ and $j$ are odd  and $\sim t^{-3}$ otherwise) and
$T>0$ ($\sim t^{-3}$ for all $i$ and $j$). \cite{GC80}
A boundary-to-bulk crossover can be observed for sites near the end of a
semi-infinite chain at $T = \infty$. \cite{SVM92}
\par
The correlation functions $\langle S_i^x (t)S_j^x \rangle$ have a much more
complicated structure in the fermion representation.
With the fermionic identity
\begin{equation}
(-1)^{  a_k^{\dag} a_k } = (a_k^{\dag} + a_k)(a_k^{\dag} - a_k)
\label{II.6}
\end{equation}
applied to (\ref{II.2}), this correlation function may be expressed
in terms of the auxiliary operators $A_k = a_k^{\dag} + a_k$ and
$B_k = a_k^{\dag} - a_k$ as follows:
\begin{equation}
\langle S_i^x (t) S_j^x\rangle = \frac{1}{4}
\langle A_1(t) B_1(t) A_2(t) B_2(t) ... A_{i-1}(t)
B_{i-1}(t) A_i(t) A_1 B_1 A_2 B_2 ...A_{j-1} B_{j-1} A_j\rangle \, .
\label{II.7}
\end{equation}
This expectation value of a product of $2(i+j-1)$
fermion operators may be expanded via Wick's theorem in terms of
more elementary expectation values.
The result is most compactly expressed as a Pfaffian: \cite{note1}
\begin{displaymath}
4 \langle S_i^x(t) S_j^x\rangle \; \;  =   \; \; \; \; \; \; \; \; \; \; \; \;
\; \; \; \; \; \; \; \; \; \; \; \; \; \; \; \; \; \; \; \; \; \; \; \;
\; \; \; \; \; \; \; \; \; \; \; \; \; \; \; \; \; \; \; \; \; \; \; \;
\; \; \; \; \; \; \; \; \; \; \; \; \; \; \; \; \; \; \; \; \; \; \; \;
\end{displaymath}
\begin{equation}
 \left. \begin{array}{ccccccc}
|\langle A_1(t)B_1(t)\rangle&\langle A_1(t)A_2(t)\rangle& \cdots &
\langle A_1(t)A_1\rangle & \langle A_1(t)B_1\rangle&
 \cdots & \langle A_1(t)A_j\rangle \\
        &\langle B_1(t)A_2(t)\rangle& \cdots &
        \langle B_1(t)A_1\rangle & \langle B_1(t)B_1\rangle & \cdots &
        \langle B_1(t)A_j\rangle \\
   &   & \cdots & \cdots & \cdots & \cdots & \cdots \\
   &   &        & \cdots & \cdots & \cdots & \cdots \\
   &   &        &        & \cdots & \cdots & \cdots \\
   &   &        &        &        & \cdots & \cdots \\
   &   &        &        &        &        & \langle B_{j-1}A_j\rangle \\
   \end{array}
   \right| \, .
\label{II.8}
\end{equation}
The square of the Pfaffian is equal to the determinant of the antisymmetric
matrix with the elements of (\ref{II.8}) above the diagonal.
The matrix elements can be evaluated from the expressions \cite{CG81}
\begin{equation}
\langle A_j(t) A_l\rangle =
\frac{2}{N+1} \sum_k \sin kj \sin kl [ \cos \varepsilon_k
t - i \sin \varepsilon_k t \tanh \frac{\beta \varepsilon_k}{2}]
\label{II.9}
\end{equation}
\begin{equation}
\langle A_j(t) B_l\rangle =
\frac{2}{N+1} \sum_k \sin kj \sin kl [ i \sin \varepsilon_k
t - \cos \varepsilon_k t \tanh \frac{\beta \varepsilon_k}{2}] \, .
\label{II.10}
\end{equation}
and the relations
\begin{equation}
\langle B_j(t)B_l\rangle =
-\langle A_j(t)A_l\rangle ;\,\,\, \langle B_j(t)A_l\rangle =
-\langle A_j(t) B_l\rangle \, .
\label{II.11}
\end{equation}
All elements (\ref{II.9}) with odd $j-l$ and
all elements (\ref{II.10}) with even $j-l$ vanish.
For $t=0$ the elements (\ref{II.9}) are, in fact, zero for all $j \neq l$.
All results for the correlation functions
$\langle S_i^x (t)S_j^x \rangle$ of the $XX$
model presented in the following have been derived via numerical
evaluation of the determinant associated with (\ref{II.8}) for systems
with up to $N=10000$ sites.
The results are not subject to finite-size effects on the time intervals
shown, except where this is explicitly stated.
%
\section{Bulk regime}
\label{III}
The spin correlation function $\langle S_i^x (t)S_{i+n}^x\rangle$
of $H_{XX}$ at $T=\infty$ is surprisingly simple: a pure Gaussian for
$n=0$ and identically vanishing for $n \neq 0$: \cite{SJL75,BJ76,CP77}
\begin{equation}
\langle S_i^x(t) S_{i+n}^x\rangle = \frac{1}{4} \delta_{n,0} \exp\left[
-\frac{J^2 t^2}{4} \right] \; \; \; \, \, \,\,\,\,\,\      (T = \infty) \, .
\label{III.1}
\end{equation}
No physical argument was ever furnished to explain this peculiar decay law.
At $T=0$ that same spin correlation function exhibits a power-law decay,
of which the leading term in an asymptotic expansion, \cite{MPS83}
\begin{equation}
\langle S_i^x(t) S_{i+n}^x\rangle = \frac{1}{4}
\frac{\sqrt{2} A^2}{(n^2 - J^2 t^2 )^{1/4}}
\,\,\,\,\,\,\,\,\,\,\,\, (T=0)
\label{III.2}
\end{equation}
with $A = 2^{1/12}\exp[3\zeta^{\prime} (-1)] = 0.64500248...$,
reflects the Luttinger liquid nature of the ground state
of $H_{XX}$. Further terms of that expansion are known for general $n$
and many more for $n=0$. \cite{MPS83,MS84}
\par
Until recently it was not at all clear whether the Gaussian and power-type
decay laws persist at any finite nonzero temperatures. Then Its, Izergin,
Korepin, and Slavnov \cite{IIKS93} established on a
rigorous basis that $\langle S_i^x (t)S_{i+n}^x\rangle$
decays exponentially for $0<T<\infty$, i.e. more slowly than (\ref{III.1})
yet more rapidly than (\ref{III.2}).
{}From the solutions of the completely integrable discrete nonlinear
Schr\"odinger model, which is related to $H_{XX}$, they were able to derive
the following expression for the two-spin correlation function:
\begin{equation}
\langle S_i^x(t)S_{i+n}^x\rangle \; \; \propto \; \; \left\{
\begin{array}{cc}
\exp f(n,0) \, , \; \; & n/Jt > 1 \\
t^{4 \nu^2} \exp f(n,t) \, , & n/Jt < 1
\end{array}
\right. .
\label{III.3}
\end{equation}
with
\begin{equation}
f(n,t) = \frac{1}{2 \pi} \int_{-\pi}^{\pi} dp \, |n - Jt \sin p| \,
 \ln |\tanh
\frac{J \cos p}{2T}| \, ,
\label{III.4}
\end{equation}
\begin{equation}
\nu =  \frac{1}{2 \pi} \ln |\tanh \frac{J \sqrt{1-(n/Jt)^2}}{2T}| \, ,
\label{III.5}
\end{equation}
valid in the space-like $(n/Jt > 1)$ or the time-like $(n/Jt < 1)$ sector
of the long-time $(Jt \rightarrow \infty)$ and/or long-distance
$(n \rightarrow \infty)$ asymptotic regime.
The function $f(n,t)$ is negative and monotonically decreasing with
increasing $T$; it diverges logarithmically at $T=\infty$, thus signalling
the change in decay law.
\par
In the high-temperature regime, the result (\ref{III.3}) for the
autocorrelation function $(n=0)$ can be brought into the more explicit form
\begin{equation}
\langle S_i^x(t) S_i^x\rangle \sim \exp\left[ - \frac{2Jt}{\pi}(1+\ln
\frac{2T}{J}) \right] \,\,\,\,\,\,\,\,\,\,\,\,\, (1 \ll T/J < \infty) \, .
\label{III.6}
\end{equation}
\par
What remains to be filled in for the bulk spin correlation functions
$\langle S_i^x (t)S_{i+n}^x\rangle$ is to connect the exponential
decay to the Gaussian
decay $(n=0)$ or the identically vanishing result $(n \neq 0)$ in the
high-temperature limit, and to the power-law decay in the low-temperature
limit.
These connections are realized by crossovers between different decay laws
at short and long times and can be investigated systematically in the data
of finite systems.
The salient features of the crossovers are described in Figs. \ref{1} and
\ref{2} for autocorrelations $(n=0)$ and in Fig. \ref{3} for
paircorrelations $(n \neq 0)$.
%
\subsection{Autocorrelations}
\label{A}
The modulus-squared spin
autocorrelation function $|\langle S_i^x (t)S_i^x\rangle|^2$
is plotted logarithmically in Fig. \ref{1} for six values of $T/J$, all
in the high-temperature regime (solid lines).
The bulk character of these results for site $i=49$ of a chain
with $N=100$ spins has been ascertained by comparison with the results of
longer chains (with up to $N=1000$ spins).
The (parabolic) dashed line represents the Gaussian (\ref{III.1}) --
the exact result for $T=\infty$.
\par
We observe that the Gaussian behavior persists at finite $T$ over some
range of short times. That range shrinks with decreasing temperature.
{}From the common short-time parabolic shape the individual lines take off
like a bundle of tangents, which represent the exponential character of the
long-time decay. The crossover takes place quite suddenly. \cite{M80}
The slight wiggles in the high-temperature data will turn into stronger
oscillations in the low-temperature regime as we shall see.
The observed decay rate in the exponential regime as represented by the
slope of the tangent lines decreases monotonically as the temperature
is lowered.
It is well matched in each case by the slope of the adjacent dot-dashed line,
which represents the decay rate
\begin{equation}
\frac{1}{\tau} = \frac{2J}{\pi}\left[ 1 + \ln \frac{2T}{J} \right]
\,\,\,\,\,\,\,\,\,\,\, (T/J \gg 1)
\label{III.7}
\end{equation}
of the asymptotic result (\ref{III.6}).
\par
The solid lines in Fig. \ref{2} show the same quantity as in Fig. \ref{1}
but now for three values of $T/J$ in the low-temperature regime.
Here the short-time Gaussian behavior has disappeared from the scene.
The steepest curve corresponds to the highest temperature $(2T/J=1.0)$.
The exponential nature of the decay (with mild oscillations superimposed)
is now realized even at relatively short times.
The average slope is perfectly consistent with the decay rate inferred from
the asymptotic result (\ref{III.3}), represented by the slope of the adjacent
dot-dashed line.
\par
At lower temperatures a crossover between exponential decay and power-law
decay makes its appearance. The power-law behavior is first seen at
short times.
In the center curve of Fig. \ref{2}, the crossover takes place prior
to $J t = 20$.
At longer times the exponential decay is still clearly visible, and the
rate of decay agrees well with the asymptotic rate
\begin{equation}
\frac{1}{\tau} = T \left[ \frac{\pi}{2} - \frac{4}{\pi} e^{-J/T} + \ldots
\, \right] \,\,\,\,\,\,\,\,\, (T/J \ll 1)
\label{III.8}
\end{equation}
extracted from (\ref{III.3}) (see adjacent dot-dashed line).  The top
curve in Fig. \ref{2} represents the $T=0$ result, which has been
investigated in previous studies.  The dashed line shows the
asymptotic power law (\ref{III.2}) for $n=0$, which matches the data
shown here extremely well (except for the oscillations).
\par
The onset of finite-size effects at longer times
is shown in the inset to Fig. \ref{2} for two cases.
The rebound of the $T=0$ correlation function at $J t \simeq 100$ can be
interpreted as the
echo from open ends of the ballistically propagating fermions.
In the $T>0$ case, the peak at $J t \simeq 100$ is absent because of
destructive interference. The first echo now occurs at $J t \simeq 200$,
the time it takes a pulse to move through the system twice.
The speed of propagation which determines the echo time is given by the
maximum fermion velocity $v$ = max$[d\epsilon_k /dk]$. For the
dispersion (\ref{II.5}) of $H_{XX}$ we have $v=J$.  In the presence of
anisotropy as realized in the $H_{XY}$ with $J_x \neq
J_y$, $\epsilon_k$ acquires a gap, $v$ decreases, and finite-size
effects set in later. \cite{ANJS} These echo effects occur at all
temperatures. They are easy to recognize in
all data produced for this study.
%
\subsection{Paircorrelations}
\label{B}
In Fig. \ref{3} we show a logarithmic
plot of $|\langle S_i^x (t)S_{i+n}^x\rangle|^2$
for $n$=4, 9, 14, 19 at the intermediate temperature $2T/J=1$.
We observe that each function is almost perfectly constant up to a time
$Jt_n \simeq n$, where it bends smoothly into exponential decay with
superimposed oscillations.
The decay time does not show any significant dependence on $n$.
The inverse decay time predicted by (\ref{III.3}) for the asymptotic
regime of the uppermost curve is given by the slope of the dashed line
and matches our data very well.
A numerical analysis of (\ref{III.3}) shows that the asymptotic decay time
increases slightly with increasing $n$ at fixed temperature.
The linear variation with $n$ of the intercepts at $Jt=0$ in this
logarithmic plot reflects the well-established exponential decay of the
equal-time correlation function
$\langle S_i^x S_{i+n}^x \rangle \sim \exp[-n/\xi(T)]$.
\par
The inset to Fig. \ref{3} shows again the curve $n=19$ of the main plot
along with curves for the same correlation function at different temperatures.
Now the crossover between the space-like and the time-like regime occurs
at one common value of $Jt$.
In the time-like regime, the slope changes from one curve to the next, which
reflects the $T$-dependence of the decay time, while the variable intercept
in the space-like regime reflects the $T$-dependence of the correlation
length.
\par
The correlation length $\xi(T)$ is known to diverge algebraically,
$\sim 1/T$, at $T=0$ and to vanish logarithmically, $ \sim 1/\ln(T)$,
at $T=\infty$.\cite{M68,BM71,T81}
We have noted in Sec. \ref{A} that the decay time $\tau (T)$ also goes to zero
logarithmically at $T=\infty$ and exhibits the same power-law divergence at
$T=0$.
Figure \ref{4} shows both the inverse correlation length and the inverse
decay time plotted versus temperature. In the $XXZ$ model, the two quantities
are expected to have more distinct temperature dependences.
The numerical results of Ref. \onlinecite{BVSM94} indicate that $\tau(T)$
stays nonzero at $T=\infty$, whereas $\xi(T)$ is expected to vanish in that
limit as it does in the $XX$ model.
%
\section{Boundary effects}
\label{IV}
Here we investigate spin autocorrelation functions
$\langle S_i^x(t) S_i^x\rangle$ of a semi-infinite
chain for sites $i = 1, 2, \ldots$
beginning at the boundary.
The actual calculations are performed for sites $i$ near one end of a finite
open chain (\ref{II.3}) of $N$ spins.
However, none of the results presented are affected by the far end
of the chain.
\par
The long-time asymptotic decay of $\langle S_i^x(t) S_i^x\rangle$ in the
boundary regime of a semi-infinite chain is different at
zero, finite nonzero, and infinite temperatures.
It is fastest at $T=\infty$ and slowest at $T=0$, like in the bulk regime,
but instead of seeing
transitions from Gaussian to exponential to power-law decay, we now
observe transitions between three types of power-law decay.
\par
At infinite temperature, the power-law long-time asymptotic decay has
the form
\begin{equation}
\langle S_i^x(t) S_i^x\rangle \sim t^{-3/2 - (i-1)(i+1)} \,\,\,\,\,\,\,\,\,\,\,
(T = \infty)
\label{IV.1}
\end{equation}
with a site-dependent exponent.
This result was inferred from an exact calculation for sites
$i = 1, 2, \ldots, 5$, and presumably holds for all sites of a semi-infinite
chain. \cite{SVM92}
The power-law decay at zero temperature as obtained from a perturbational
treatment of the Pfaffian (\ref{II.8}) is site-independent,\cite{PM78}
\begin{equation}
\langle S_i^x(t) S_i^x\rangle \sim t^{-1} \,\,\,\,\,\,\,\,\,\,\,\,\,
(T = 0) \, ,
\label{IV.2}
\end{equation}
and different from the $t^{-1/2}$ decay in the bulk regime.
Our numerical analysis of the same determinantal expression at finite nonzero
temperatures strongly indicates that the long-time asymptotic decay is a
power-law with yet a different site-independent exponent:
\begin{equation}
\langle S_i^x(t) S_i^x\rangle \sim t^{-3/2} \,\,\,\,\,\,\,\,\,\,\,\,\,\,\,
(0 < T < \infty) \, .
\label{IV.3}
\end{equation}
\par
An analytic calculation, which confirms this decay law, is presented in
Appendix A.
The five curves in Fig. \ref{5} represent the quantity
$|\langle S_i^x(t) S_i^x\rangle |^2$ for $i=2$
at various temperatures in a log-log plot.
The data are subject to strong oscillations, which makes it hard to
distinguish the different decay laws in this graphical
representation. Therefore we have smoothed the data at $Jt > 40$ as
described in the caption.
The three different slopes of the dashed lines represent the decay laws
(\ref{IV.1}) - (\ref{IV.3}), i.e. $t^{-9/2}$ for $i=2$ (bottom line),
$t^{-3/2}$ (intermediate lines), and $t^{-1}$ (top line).
\par
The decay laws of $\langle S_i^x(t) S_i^x\rangle$
at a given temperature and for a given
site on the semi-infinite chain may undergo one or several crossovers.
Here we have to deal with bulk-to-boundary crossovers in addition to
crossovers between different temperature regimes.
We first look at the two types individually and then in combination.
\par
At $T=\infty$ we observe a bulk-to-boundary crossover, i.e. a crossover
from Gaussian decay (\ref{III.1}) at short times to power-law decay
(\ref{IV.1}) at long times. This is illustrated in Fig. \ref{6} for the sites
$i = 2, \ldots, 11$ near the end of a semi-infinite chain.
The time $Jt_c$ marking the onset of the crossover depends linearly on
the distance of the site $i$ from the boundary as shown in the inset to
Fig. \ref{5}.
A corresponding bulk-to-boundary crossover between the power laws
$t^{-1/2}$ and $t^{-1}$
takes place in $\langle S_i^x(t) S_i^x\rangle$ at $T=0$.
\par
Now we keep the site fixed at $i=2$ close to the boundary and vary the
temperature.
At high-temperatures $(T/J \gg 1)$ we can observe an infinite-to-finite-$T$
crossover similar to the one portrayed in Fig. \ref{1} for the bulk regime.
But here in the boundary regime, it is a crossover between two power laws:
$t^{-9/2}$ and $t^{-3/2}$.
This crossover is illustrated in Fig. \ref{7}.
The solid lines represent smoothed data at $2T/J = 10^3$ (top) and
$2T/J =10^5$ (bottom) in a log-log plot. It shows how the crossover is shifted
to
longer times as the temperature approaches infinity.
A zero-to-nonzero-$T$ crossover between the respective
power-laws $t^{-1}$ and $t^{-3/2}$ can be observed at low temperatures
$(T/J \ll 1)$.
This is illustrated in the inset to Fig. \ref{7}. Again the crossover is
different from the corresponding (power-law-to-exponential) crossover
in the bulk regime (Fig. \ref{2}).
\par
The two types of
crossover in $\langle S_i^x(t) S_i^x\rangle$ of the semi-infinite
$XX$ chain, which we have described separately, namely the bulk-to-boundary
crossover (Fig. \ref{6}) and the infinite-to-finite-$T$ or zero-to-nonzero-$T$
crossover (Fig. \ref{7}), may actually occur in one and the same data set.
One case in point is demonstrated in Fig. \ref{8}.
It shows a logarithmic plot
of $|\langle S_i^x(t) S_i^x\rangle |^2$ at high temperatures
$(T/J \gg 1)$ for a site of the infinite chain (solid lines) and a site not
too far from the end of a semi-infinite chain (dashed lines).
The solid lines, which represent data already shown in Fig. \ref{1},
exhibit the familiar infinite-to-finite-$T$ crossover pertaining to the bulk
regime.
The same crossover between decay laws (\ref{III.1}) and (\ref{III.6}) is
also observable in the dashed lines, but here it is followed by the
bulk-to-boundary crossover at finite $T$, i.e. between decay laws
(\ref{III.6}) and (\ref{IV.3}).
\par
Note that in the lowest dashed curve, which corresponds to the highest
temperature, the intermediate decay law has virtually disappeared.
Here the two crossovers overlap.
At still higher temperatures their order is reversed.
The Gaussian decay
law (\ref{III.1}) crosses over (bulk-to-boundary) to the power-law
(\ref{IV.1}), which in turn crosses over (infinite-to-finite-$T$) to the
power law (\ref{IV.3}).
Corresponding crossover combinations take place in the low-temperature
regime.
\acknowledgments
The work at URI was supported by the U. S. National Science Foundation,
Grant DMR-93-12252 and by the NCSA Urbana Champaign.
J.S. gratefully acknowledges the generous hospitality of the Department of
Physics, University of Rhode Island, and the financial support of the
Max Kade Foundation during the time when this work was begun. We thank
L.L. Gon\c{c}alves for useful comments.
\pagebreak
\appendix
\section{Decay Laws in the Boundary Regime}
The $t^{-3/2}$ decay law of the spin autocorrelation function
$\langle S_i^x(t) S_i^x \rangle$ in the boundary regime at finite nonzero $T$
can be recovered by the following analysis of the Pfaffian (\ref{II.8})
and its elements.
For $N \rightarrow \infty$ all non-vanishing expectation values
(\ref{II.9}) and (\ref{II.10}) are equal to
\begin{equation}
f_{jl}(t) = \frac{4}{\pi} (-)^{j-l} \int_{-1}^{1} dx \; \sqrt{1 -
x^2} \;
U_{j-1}(x) \; U_{l-1}(x)\;  e^{i J x t} \; f(\beta J x)  \label{A.1},
\end{equation}
where $f(x) = (e^x + 1)^{-1}$ is the Fermi function and $U_{n}(x)$
is a Chebyshev polynomial of the second kind.
The long-time behavior of $f_{jl}(t)$ is determined by the singularities
of its Fourier transform
\begin{equation}
\phi_{j l}(x) = \sqrt{1-x^2} \; U_{j-1} \; (x) U_{l-1}(x) \; f(\beta
J x) \;
\Theta(1-x^2)
\label{A.3}.
\end{equation}
At nonzero $T$ the only singularities of (\ref{A.3}) are the square-root
cusps $\sqrt{1-|x|} \,  \Theta(1-|x|)$
at $x \rightarrow \pm 1$, which yield
$f_{j l}(t) \sim e^{\pm i J t} t^{- 3/2}$.
At $T=0$ an additional singularity in $\phi_{j l}(x)
$ at $x=0$ is generated by the discontinuity in the Fermi function,
but it contributes only to leading order if both polynomials in
(\ref{A.3}) are even, i.e. for odd $j$ and $l$:
$f_{j l}(t) \sim t^{- 1}$.
\par
Instead of the Pfaffian (\ref{II.8})
we study the associated antisymmetric
$(4 i - 2) \times (4 i - 2)$ matrix. Its determinant is equal
to the square of (\ref{II.8}) for $i=j$. That matrix naturally divides into
four $(2 i - 1) \times (2 i - 1)$ blocks:
\begin{equation}
16 \langle S_i^x(t) S_i^x \rangle^2 =
 \left|
\begin{array}{cc}
 {\bf M} &  {\bf A}   \\
 {\bf A}  & \ {\bf M^\prime}
\end{array}
\right| \, .
\label{A.8}
\end{equation}
The blocks ${\bf M}$ and ${\bf M^\prime}$ contain only
time-independent elements, $f_{j l}(0)$, and the block ${\bf A}$ only
time-dependent elements, $f_{j l}(t)$.
Block ${\bf M}$ turns out to have the following general structure:
\begin{equation}
{\bf M} =
\left(
       \begin{array}{cccccccccc}
0 & 0 & 0 & f_{12}(0) & 0 & 0 & 0 & f_{14}(0) & 0 & \cdots \\
0 & 0 & -f_{12}(0) & 0 & 0 & 0 & -f_{14}(0) & 0 & 0 &\cdots \\
0 & f_{12}(0) & 0 & 0 & 0 & f_{23}(0) & 0 & 0 & 0 &  \cdots \\
-f_{12}(0) & 0 & 0 & 0 & -f_{23}(0) & 0 & 0 & 0 & -f_{25}(0) &  \cdots \\
0 & 0 & 0 & f_{23}(0) & 0 & 0 & 0 & f_{34}(0) & 0 &  \cdots \\
0 & 0 & -f_{23}(0) & 0 & 0 & 0 & -f_{34}(0) & 0 & 0 & \cdots \\
0 & f_{14}(0) & 0 & 0 & 0 & f_{34}(0) & 0 & 0 & 0 &  \cdots \\
-f_{14}(0) & 0 & 0 & 0 & -f_{34}(0) & 0 & 0 & 0 & -f_{45}(0) &  \cdots \\
0 & 0 & 0 & f_{25}(0) & 0 & 0 & 0 & f_{45}(0) & 0 & \cdots \\
\cdots & \cdots & \cdots & \cdots & \cdots & \cdots & \cdots & \cdots
& \cdots & \cdots
       \end{array}
\right)
\label{A.10}.
\end{equation}
Block ${\bf M^{\prime}}$ has a similar distribution of nonzero
elements.
The leading long-time term in the expansion of (\ref{A.8}) is the one
which contains the smallest possible number of time-dependent elements,
i.e. the largest possible number of elements from
${\bf M}$ and ${\bf M^{\prime}}$.
Given the structure of ${\bf M}$, it is not possible to pick more than
$2i-2$ elements from ${\bf M}$ in the expansion of (\ref{A.8}).
The same holds true for ${\bf M^{\prime}}$.
Therefore, the leading term in (\ref{A.8}) contains exactly two
time-dependent elements $f_{j l}(t)$ from ${\bf A}$, which explains the
decay laws (\ref{IV.2}) and (\ref{IV.3}) at $T=0$ and $0<T<\infty$,
respectively.
At $T = \infty$ all elements of ${\bf M}$ and ${\bf M^{\prime}}$ vanish,
and the asymptotic long-time behavior is solely
determined by block ${\bf A}$.
This leads to the site-dependent decay law (\ref{IV.1}) as explained in
Ref. \onlinecite{SVM92} \cite{zcorr}.
\pagebreak

\pagebreak

\begin{figure}
\caption[eins]{
Spin autocorrelation function $\langle S_i^x(t)S_i^x\rangle$ in
the bulk regime of $H_{XX}$
at high temperatures. Plotted is the logarithm of
$| \langle S_i^x(t)S_i^x\rangle |^2 $ evaluated at increments $Jdt$ = 0.4
for site $i=49$ in an open chain of $N$=100 spins, for $2T/J = 10, 10^2,
\ldots, 10^6$ (solid lines, top to bottom).
The exact result (\ref{III.1}) for $T=\infty$ is shown as dashed line.
The slopes of the dot-dashed straight-line segments represent the decay rate
of (\ref{III.6}).
}
\label{1}
\end{figure}

\begin{figure}
\caption[zwei]{
Spin autocorrelation function $\langle S_i^x(t)S_i^x\rangle$ in the
bulk regime of $H_{XX}$ at low temperatures. Plotted is the logarithm
of $| \langle S_i^x(t)S_i^x\rangle |^2$ evaluated at increments $Jdt =
0.4 $ for site $i=49$ in an open chain of $N=100$ spins at $2T/J = 1.0,
0.1$ and for site $i=249$ in a system of size $N=500$  at $T=0$
(solid lines, bottom to top).
The dashed line represents (\ref{III.2}).
The slopes of the dot-dashed straight-line
segments represent the decay rate inferred from the asymptotic
expression (\ref{III.3}).  The inset shows the results at $2T/J=0$
(upper curve) and 0.1 (lower curve) again over a longer time interval
and (now in both cases) for $N=100$
in order to illustrate the onset of finite-size effects in the form of
echos due to ballistically propagating pulses reflected at the open ends of
the chain.
}
\label{2}
\end{figure}

\begin{figure}
\caption[drei]{
Spin paircorrelation function $\langle S_i^x(t)S_j^x\rangle$
in the bulk regime of $H_{XX}$
at $2T/J=1$, for $j=49$ and $i=30, 35, 40,$ and $45$
(solid lines, top to bottom) in an open chain with $N=100$ spins.
Plotted is the logarithm of $|\langle S_i^x(t)S_j^x\rangle|^2$
evaluated at increments $Jdt = 0.4$
The inset shows the same function in the same representation, but here the
sites are kept fixed $(i=30 , j=49)$ and the temperature is varied
$(2T/J=0.1, 1, 10)$.
The slopes of the dot-dashed straight-line segments represent the decay rate
inferred from the asymptotic expression (\ref{III.3}).
}
\label{3}
\end{figure}

\begin{figure}
\caption[vier]{Temperature dependence of the inverse decay time
$1/\tau$ as derived from
the asymptotic expression (\ref{III.3}) (solid line) and from our numerical
analysis of large but finite chains (circles).
The numerical data were derived from an exponential fit to $|\langle
S_i^x(t)S_i^x\rangle|^2$ for $i=49$ in an open chain of $N=100$ spins,
evaluated for $20 \leq J t \leq 60$ at increments $J d t = 0.4$.
The dashed lines represent the approximations (\ref{III.7}) and
(\ref{III.8}) to (\ref{III.3}) for high and low $T$, respectively. For
comparison we also show (dot-dashed) the temperature dependence of the
inverse correlation length $1 / \xi$ as given by (\ref{III.3}). Data
read off Fig. 2 of Ref. \onlinecite{T81} coincide with the curve
shown, as do data derived from our own numerical results for $\langle
S_i^x S_j^x \rangle / \langle S_{i+1}^x S_j^x \rangle $ with $|i-j| =
10$ or $20$ in an open chain of $N=100$ spins.
}
 \label{4}
 \end{figure}

\begin{figure}
\caption[fuenf]{
Spin autocorrelation function $\langle S_i^x(t)S_i^x\rangle$ in
the boundary regime of
$H_{XX}$ at $2T/J = 0, 0.1, 1, 10, \infty$ (solid lines, top to bottom).
Shown is the
quantity $|\langle S_i^x(t)S_i^x\rangle |^2 $ evaluated at increments
$Jdt = 0.4 \, (J t \leq 40 ) , Jdt = 1 \, (40 \leq J t \leq 100)$, and $J d
t = 10 \, (100 \leq J t \leq 2000)$ in a doubly logarithmic plot.
The data pertain to the site $i=2$ in an open chain of $N=10000$ spins.
At times $Jt > 40$ most of the data were smoothed by the following two-step
procedure:
(i) retain all local maxima of the data set;
(ii) eliminate all local minima from the remaining data set. The $2 T
/ J = 0.1 $ data did not require smoothing.
The dashed lines represent the power laws $t^{-1}$ (top), $t^{-3/2}$
(center), and $t^{-9/2}$ (bottom).
}
\label{5}
\end{figure}

\begin{figure}
\caption[sechs]{
Logarithmic plot of the square of the spin autocorrelation function
$\langle S_i^x(t)S_i^x\rangle$ in the boundary regime of $H_{XX}$ at $T=\infty$
for $i=2, \ldots, 11$ (top to bottom).
The dashed line represents the Gaussian (\ref{III.1}) pertaining to the
bulk limit ($i \rightarrow \infty)$.
The data for $i \leq 5$ are obtained from the exact expressions given in
Ref. \onlinecite{SVM92} and the data for $i > 5$ from an open chain
with $N=100$ spins.
The inset shows the time $Jt_c$ at which the relative deviation of
$\langle S_i^x(t)S_i^x\rangle^2$ from the Gaussian first exceeds one percent.
}
\label{6}
\end{figure}

\begin{figure}
 \caption[sieben]{
The main plot shows smoothed data of the same quantity as in Fig. \ref{5}
at two very high temperatures $(2T/J = 10^3 , 10^5 )$ and in the inset for
two very low temperatures $2T/J = 2 \cdot 10^{-3}, 10^{-2})$. The
dashed lines in the main plot represent the power laws $t^{-3/2}$
(top) and $t^{-9/2}$ (bottom). The dashed lines in the inset represent
data for temperatures $2T / J = 0, 0.1$ (i.e. the two uppermost
solid lines of Fig. 5).
}
\label{7}
\end{figure}

\begin{figure}
\caption[acht]{
Spin autocorrelation function $\langle S_i^x(t)S_i^x\rangle$ in the bulk regime
and in
the boundary regime of $H_{XX}$ at high temperatures.
Plotted is the logarithm of $|\langle S_i^x(t)S_i^x\rangle | ^2$ evaluated at
increments
$Jdt = 0.4$ for site $i=49$ (solid lines) and site $i=11$ (dashed lines)
in an open chain of $N=100$ spins for $2T/J = 1, 10, 10^3,$ and $10^5$.
The dot-dashed line represents the Gaussian decay law (\ref{III.1}).
}
\label{8}
\end{figure}


\begin{references}
\bibitem{note0} One successful early application of this method at $T=0$
        was reported in Ref. \onlinecite{PM78}.
\bibitem{note1} References on the properties of Pfaffians and
        on their use for the calculation of dynamic correlations in $XX$
        chains can be
        found, for example, in Appendix A of Ref. \onlinecite{SVM92}.
\bibitem{zcorr} The $t^{-2}$ and $t^{-3}$ decay laws of the spin pair
correlation function $\langle S_i^z (t)S_j^z \rangle$ in the boundary
regime can be explained by observing that $4 \langle S_i^z
(t)S_j^z \rangle = (-)^{i-j}(f_{ij}(t))^2$.
\end{references}
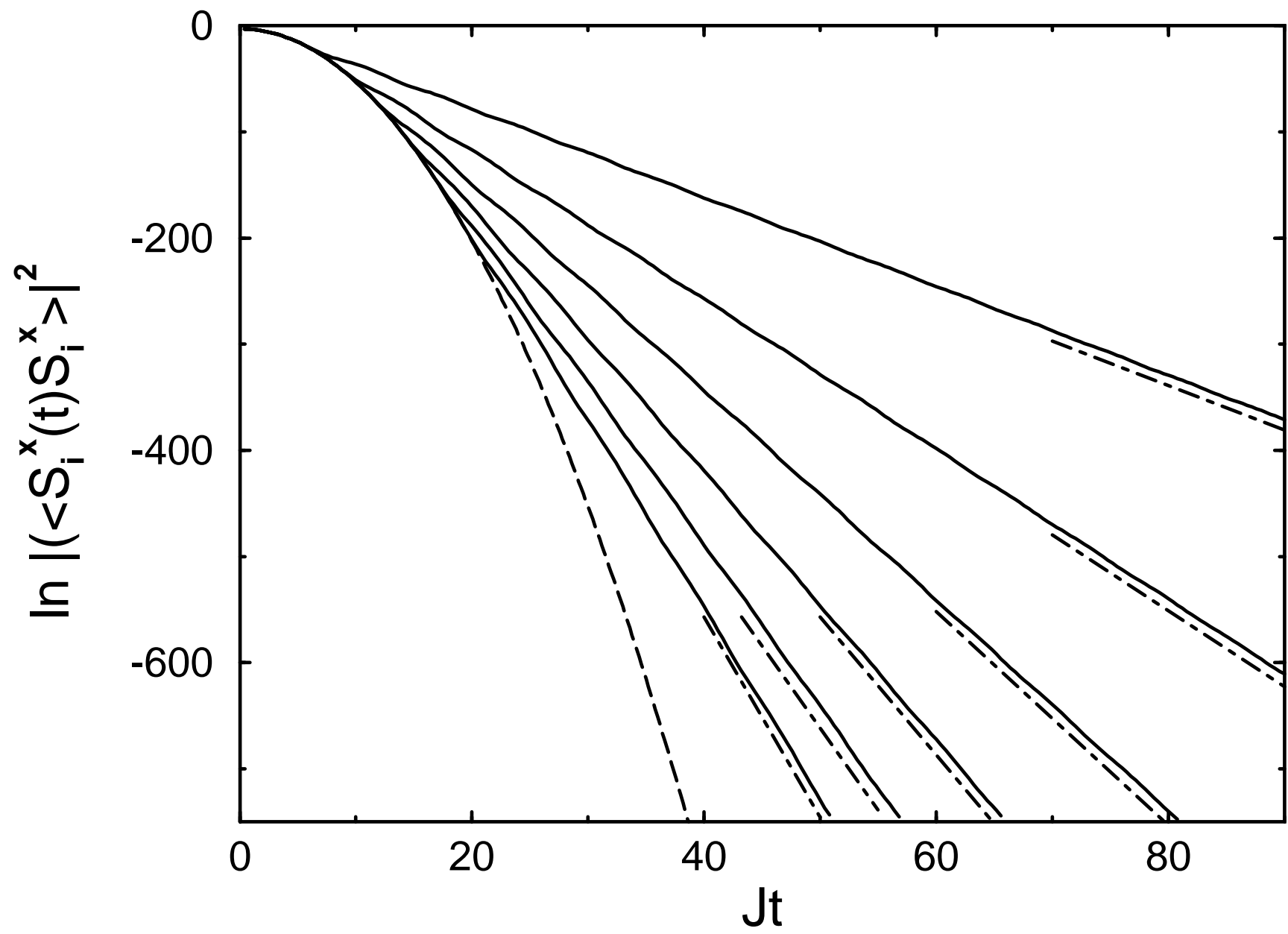
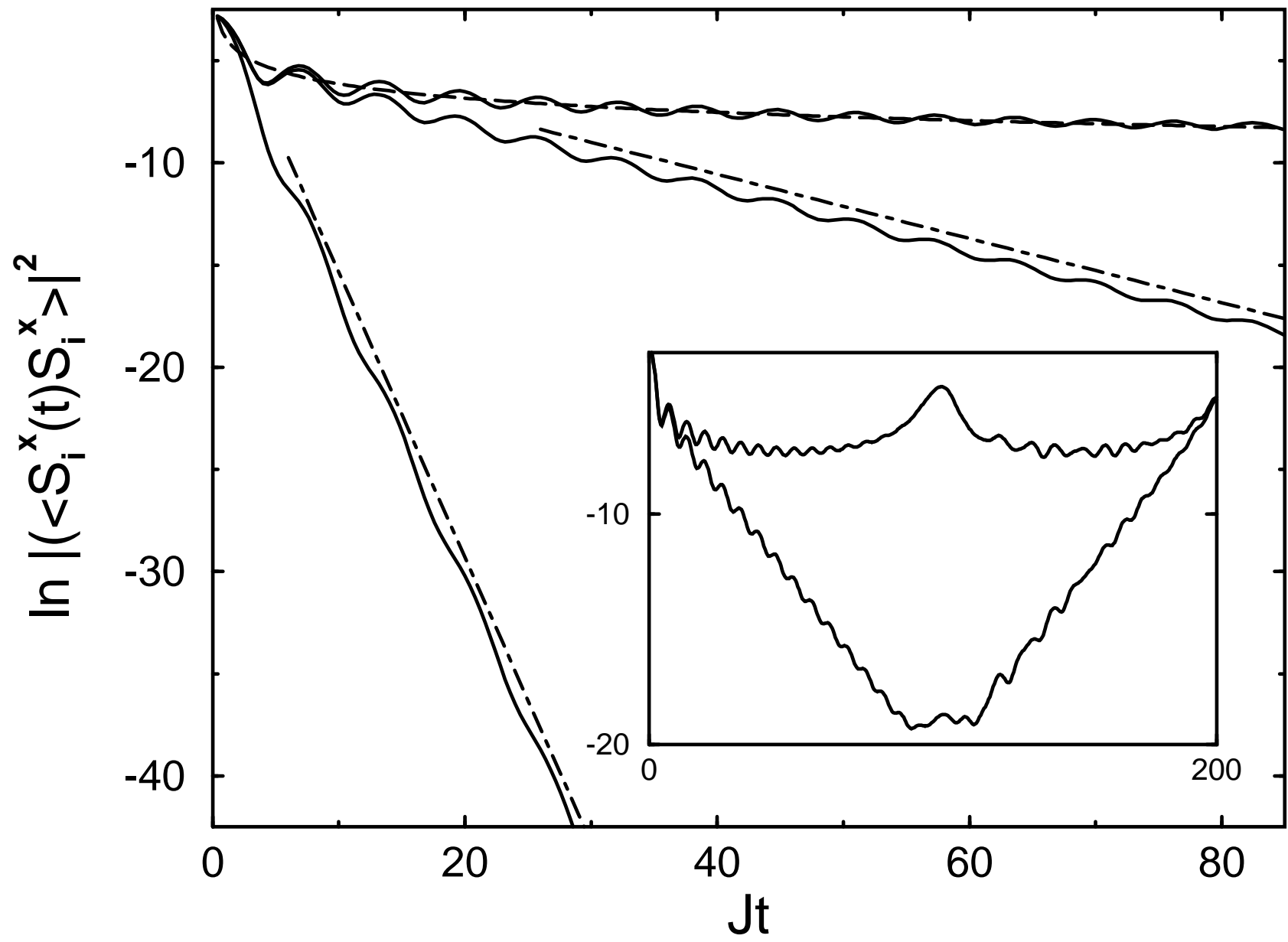
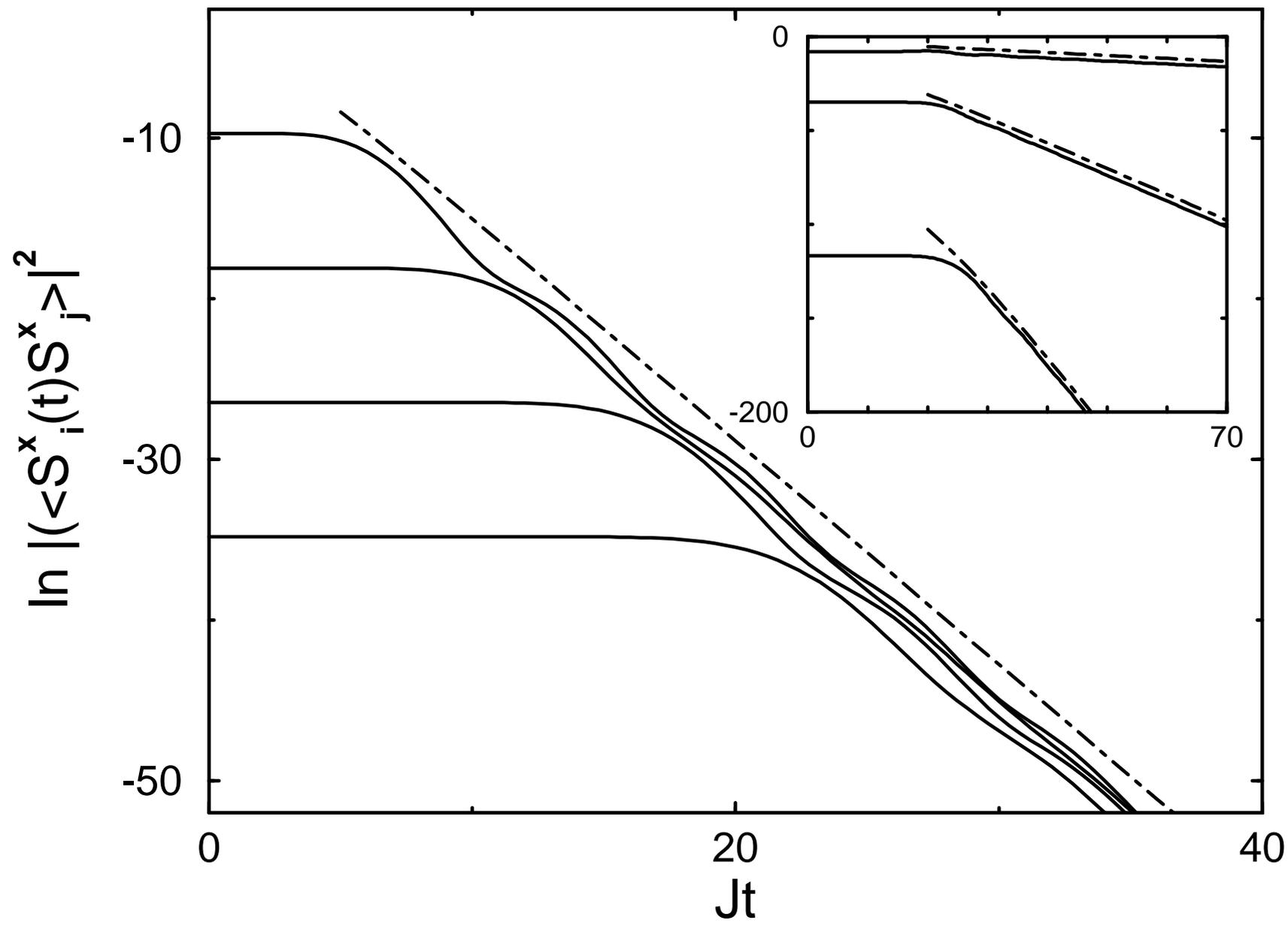
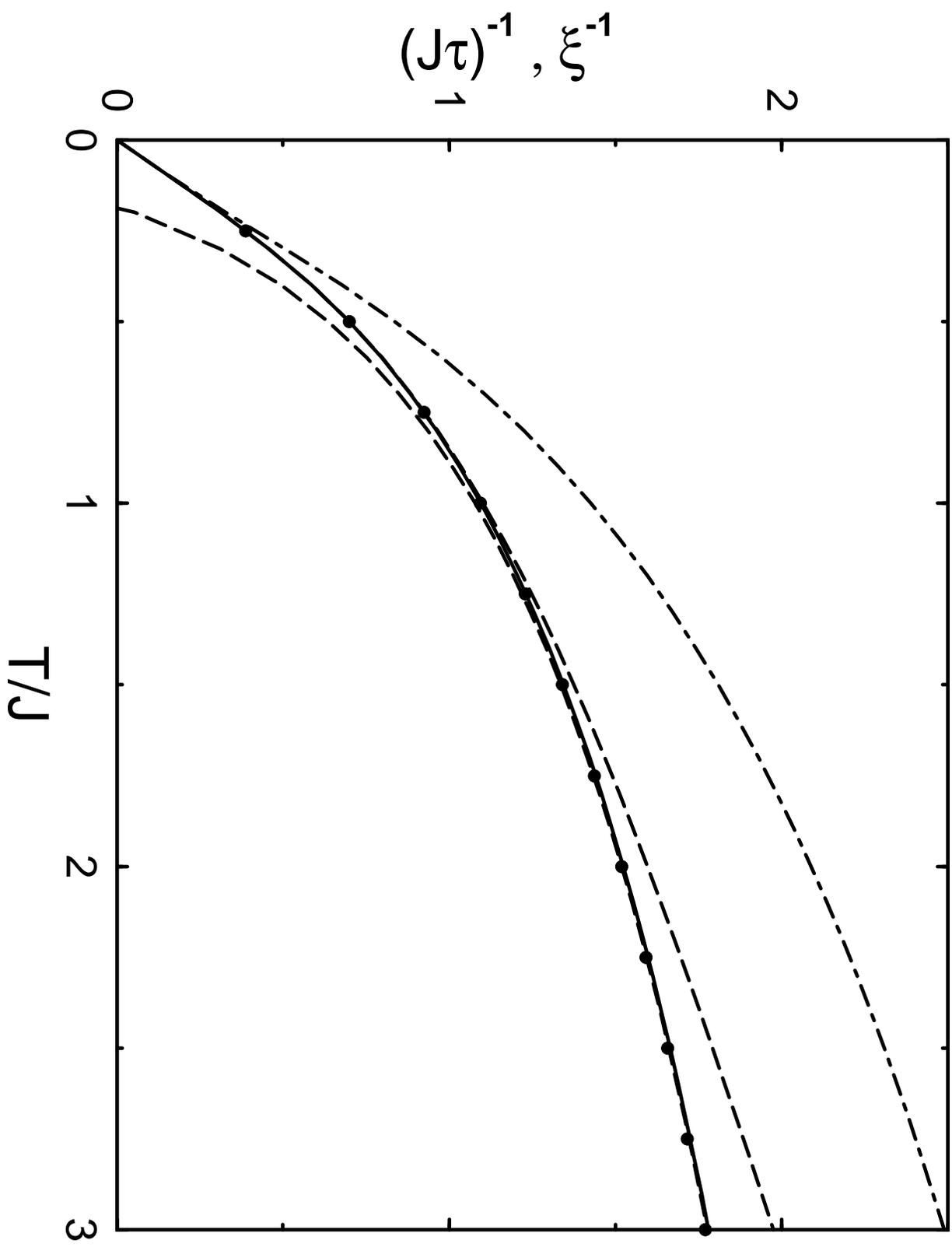
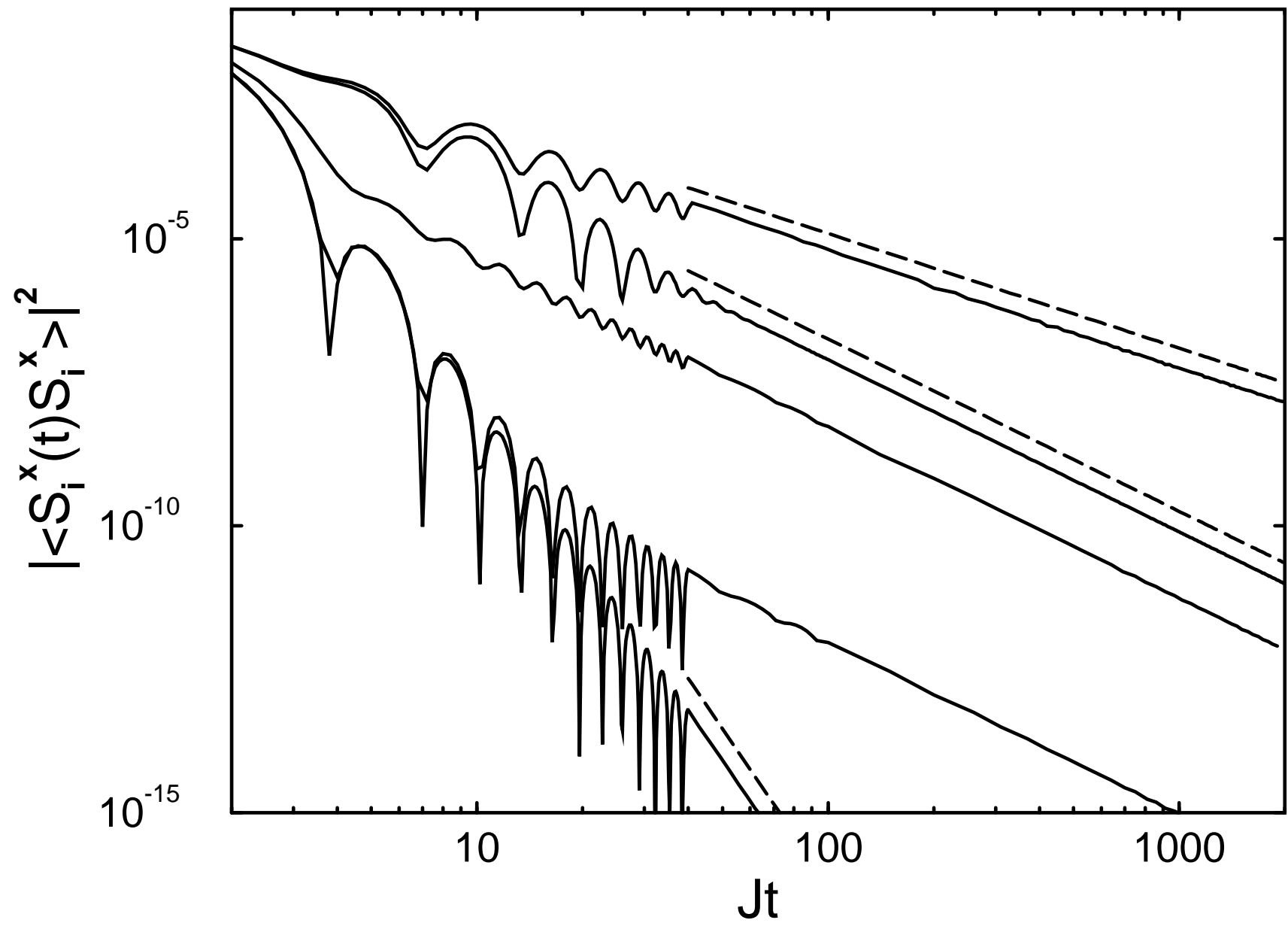
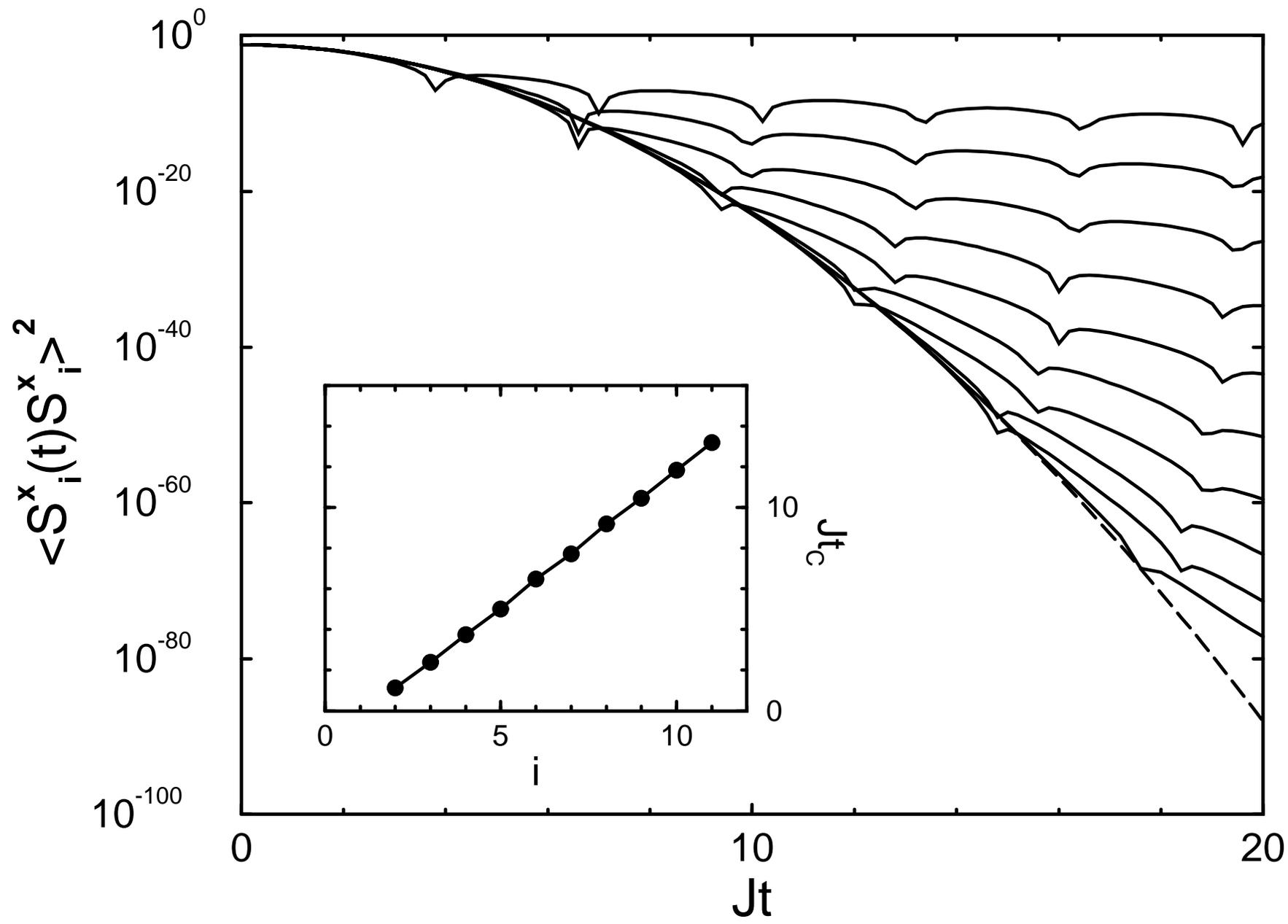
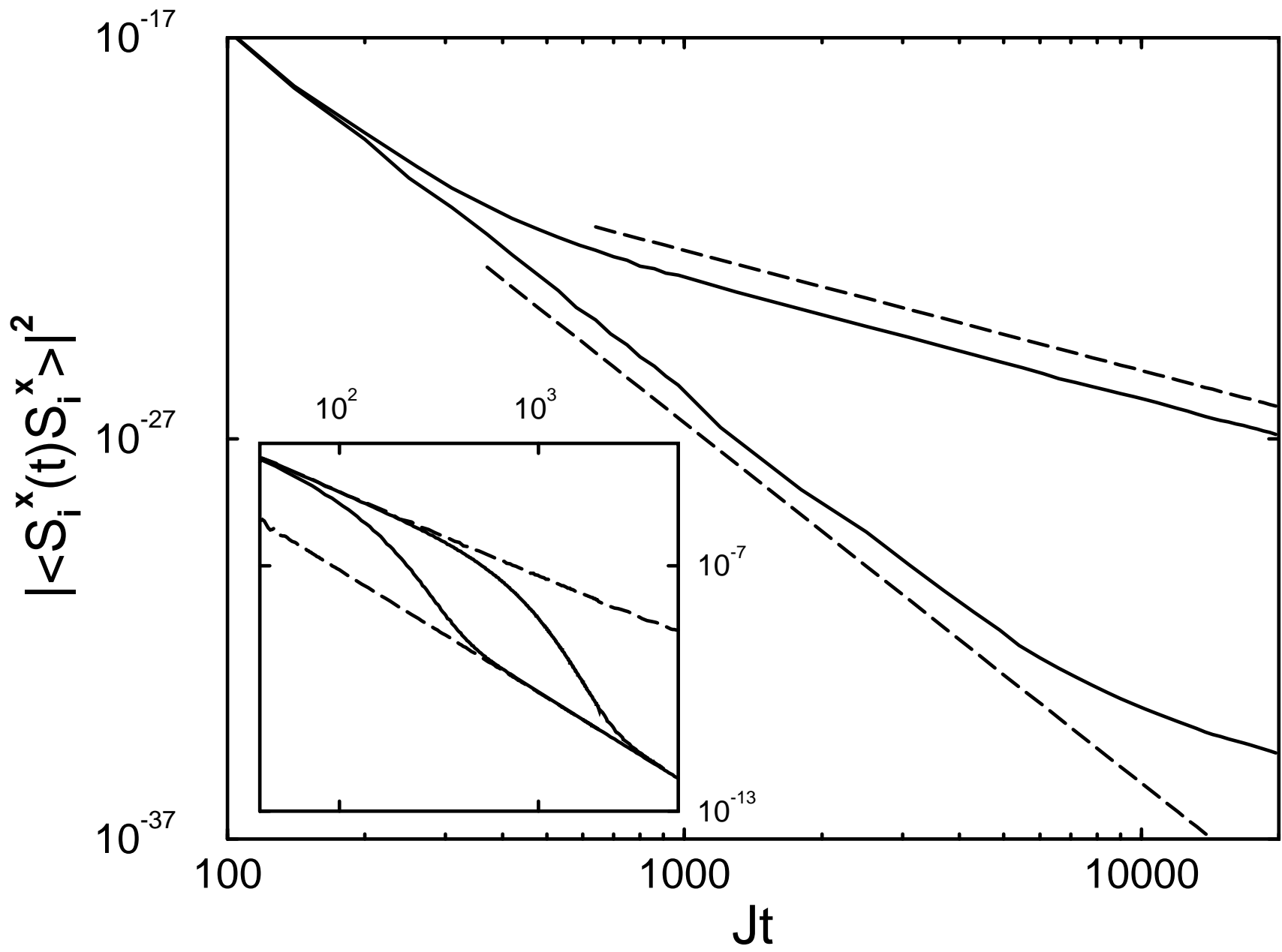
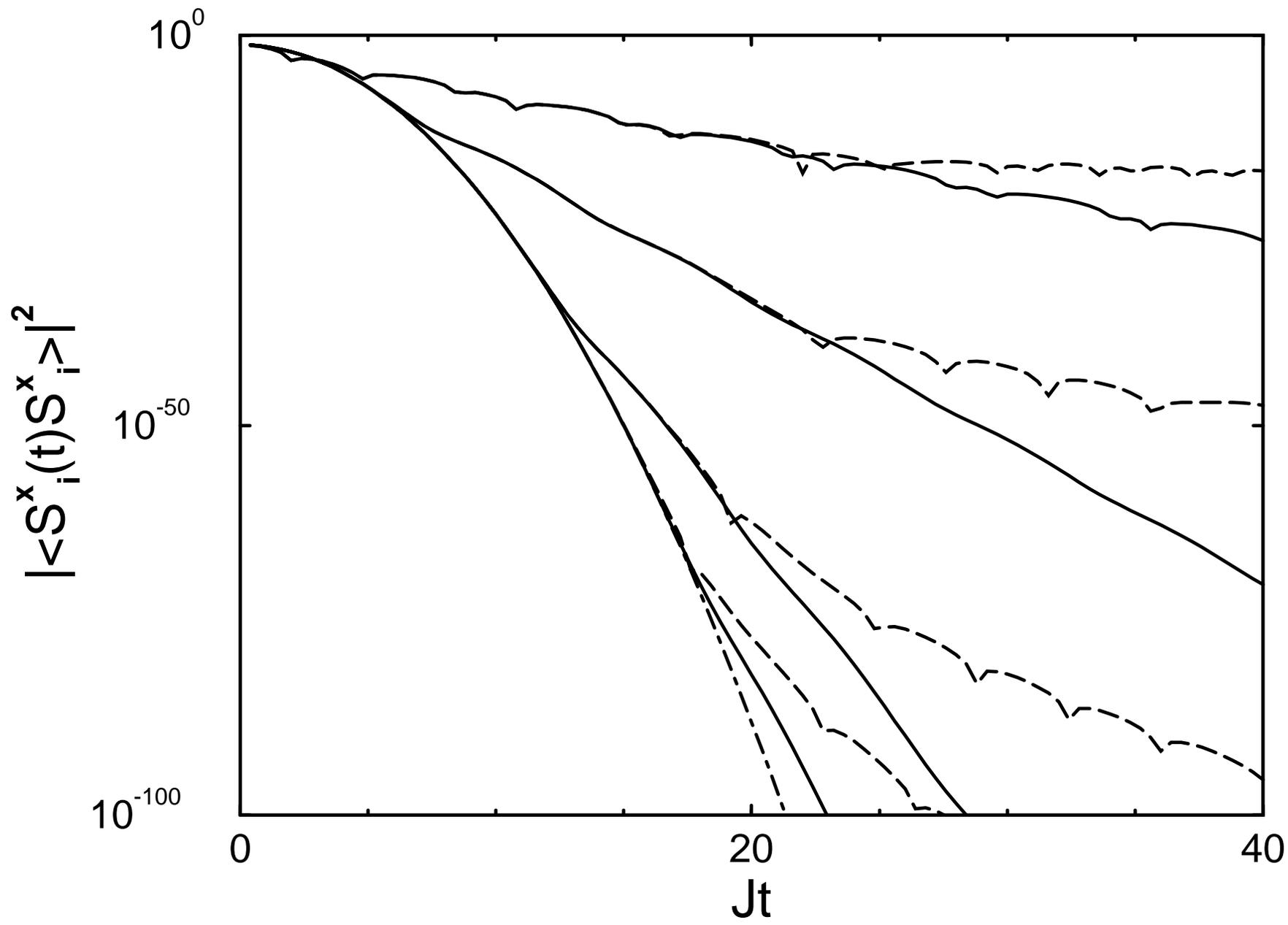
\end{document}